\documentclass[%
 aip,
 jcp,%
 amsmath,amssymb,
 reprint,%
]{revtex4-1}
\usepackage{graphicx}
\usepackage{dcolumn}
\usepackage{bm}
\usepackage{color}
\definecolor{rouge}{gray}{0}
\begin{document}

\preprint{AIP/123-QED}

\title[]{Huge Seebeck coefficients in non-aqueous electrolytes}

\author{M. Bonetti}
 \email{marco.bonetti@cea.fr}
\author{S. Nakamae}
\author{M. Roger}
\affiliation{Service de Physique de l'Etat Condens\'e, CEA-IRAMIS-SPEC
(CNRS-MPPU-URA 2464)\\
CEA-Saclay, F-91191 Gif-sur-Yvette Cedex, France.}
\author{P. Guenoun}
\affiliation{%
IRAMIS, LIONS, UMR SIS2M CEA-CNRS\\
CEA-Saclay, F-91191 Gif-sur-Yvette Cedex, France.}

\date{\today}

\begin{abstract}
The  Seebeck coefficients   of the
 non-aqueous electrolytes tetrabutylammonium 
nitrate, tetraoctylphosphonium bromide
 and tetradodecylammonium nitrate in 1-octanol, 1-dodecanol and ethylene-glycol
are measured  in a temperature range from $T$=30 to $T=45^\circ$C. 
The Seebeck coefficient is generally of the order
of a few hundreds of microvolts per Kelvin for aqueous solution of inorganic
ions. Here we report huge values of 7~mV/K at 0.1M concentration
 for tetrabutylammonium nitrate in 1-dodecanol.
 These striking  results open the 
question of unexpectedly large kosmotrope or \emph{``structure making
effects''} of tetraalkylammonium ions on the structure of alcohols.

\end{abstract}

\maketitle

\section{Introduction}
The application of a temperature difference $\Delta T$ across a solid conductor
causes mobile charge carriers to diffuse from hot to cold regions, giving rise
to a thermoelectric voltage $\Delta V=-S_e\Delta T$. 
The prefactor $S_e$ is called
the thermopower or \emph{Seebeck coefficient} since the discovery 
of this phenomenon by
 Seebeck in 1821. This property enables the conversion of heat to 
electricity. In ordinary metals like copper, $S_e$ is
of the order of 10~$\mu$V/K. In the mid 20$^{th}$ century
much higher Seebeck coefficients of a few hundreds of $\mu$V/K were obtained in
low-gap  semi-conductors, and those materials are still the object of intense
research activities, with the perspective of converting low-grade wasted
heat into electric energy.\cite{mahan,chen}
Semiconductors have reached their limit from a material 
perspective.\cite{mahan}
Research is now focussed on improving their performance by nanostructuring
existing materials, and this
represents a substantial production cost.\cite{chen}  

Thermoelectric effects also occur in liquid electrolytes, as first
observed towards the end of the 19$^{th}$ century. Electrolytes are 
characterised by the presence of several ion species. A temperature gradient
induces gradients in the concentrations of different ionic species. 
This is known as the Soret effect,\cite{soret} which couples to a Seebeck
effect in conducting fluids.
The possibility of converting heat into electricity has also been studied in
thermogalvanic cells using two identical electrodes at different 
temperatures, where chemical reactions take place.\cite{mua,hudak,nano,piazza}
 Thermogalvanic cells use  electrolytes having  large
 Seebeck coefficients (of the order of 1~mV/K in aqueous potassium 
ferrocyanide/ferricyanide solutions).\cite{mua,nano}

However the efficiency of the thermal to electrical  energy conversion is not
governed solely by a material's Seebeck coefficient. Rather it is
measured by a dimensionless number,
the ``figure of merit'' $ZT=TS_e^2(\sigma/\kappa)$, 
where $T$ is the temperature,
 and ($\sigma/\kappa$) represents the ratio of 
the electrical to the thermal conductivity. Although thermogalvanic 
cells have a high
Seebeck coefficient their relatively low electrical conductivity reduces their 
figure of merit by two orders of magnitude with respect to that of solid-state
thermoelectric devices using low-gap semiconductors.\cite{mua}

Concerning thermocells, the investigations have been  mainly restricted to
aqueous solutions of inorganic ions. Extending this field of research
 to organic electrochemistry, using organic macroions in non-aqueous
solutions, ionic liquids etc ..., 
offers a quasi-infinite number of combinations.
 We show that solutions
of tetraalkylammonium ions in organic solvents (octanol, dodecanol, 
ethylene-glycol, ...)  reach unprecedented large Seebeck coefficients 
of the order of 10~mV/K.
This result opens
the perspective of engineering  electrolytes with huge Seebeck coefficients.

 The Seebeck coefficient is closely related to the structural entropy induced
by the interactions between ions and solvent. This has been thoroughly studied
in water solutions\cite{marcus1} and in a lesser extend in non-aqueous 
solvents.\cite{marcus2} There is a pure electrostatic contribution to this
entropy coming from the interaction of the ion charge with the electric dipoles
of the solvent considered as a homogeneous dielectric medium. This 
contribution is generally evaluated through the simple Born model.\cite{born} 
There are also more subtle effects coming from the modification of hydrogen
bonds in the vicinity (first, second solvation shells and beyond...) of
the ions which can increase (kosmotrope effect) or destroy (chaotrope effect)
the local order of the surrounding solvent molecules. The
ions are named \emph{structure makers} in the first case and 
\emph{structure breakers} in the second. We show that the pure 
electrostatic contribution to the transported entropy (Born model) is small
and conjecture that the observed Seebeck coefficient arises mainly from
 kosmotrope
effects created by tetraalkylammonium ions on the local structure of alcohols.
This is a challenging open question, which needs further 
theoretical and experimental studies.

A summary of the theoretical background on the Soret and Seebeck effects
in electrolytes is reviewed in Section II. The experimental set-up is described
in section III and the results are discussed in section IV.

\section{Theoretical background}

\subsection{Particle and heat Flux}
We consider an electrolyte containing $\nu$ types of charged particles with
number densities $n_i$ and electrochemical potentials $\tilde\mu_i$. 
Slightly out
of equilibrium, the particle currents $J_{Ni}$ and the total heat flow $J_Q$
obey the linear equations:
\begin{eqnarray}
J_{Ni}&=-\frac{L_{ii}}{k_BT}\left[\nabla\tilde\mu_i+
\frac{\overline{\overline Q}_i}{T}\nabla T\right] \\
J_Q&=\sum_{i=1}^\nu \overline{\overline Q}_i J_{Ni}-\kappa\nabla T,
\end{eqnarray}
where $L_{ii}$ are Onsager coefficients,
 $\overline{\overline Q}_i$ is \emph{the heat transported}
 by particles of type $i$
and $\kappa$ is the thermal conductivity of the electrolyte. We closely 
follow the notations of Agar\cite{agarrev} and we have 
$\overline{\overline Q}_i=T\overline{\overline {\cal S}}_i$, where
$\overline{\overline {\cal S}}_i$ is named \emph{the transported entropy}.

In the presence of an electrostatic potential $V(r)$, the electro-chemical
potential is:
\begin{equation}
\tilde\mu_i=\mu_i[n_i(r),T(r)]+q_iV(r)
\end{equation}
where $q_i$ is the charge of the particles. We thus have:
\begin{equation}
\nabla\tilde\mu_i=\left.\frac{\partial\mu_i}{\partial n_i}\right|_{T=C^{st}}
\nabla n_i +\left.\frac{\partial\mu_i}{\partial T}\right|_{n_i=C^{st}}\nabla T
+q_i \nabla V
\end{equation}
From Maxwell relations:
$$
\frac{\partial\mu_i}{\partial T}=
-\frac{\partial ({\cal S}/\Omega)}{\partial n_i}=-s_i
$$
where ${\cal S}$ and $\Omega$ represent the entropy and 
 volume of the system
and $s_i$ is the partial molar entropy. At low concentration, 
assuming a non-interacting gas of particles, we have:
$$
\frac{\partial\mu_i}{\partial n_i}=\frac{k_BT}{n_i}
$$
and Eq. (1) reads:
\begin{equation}
J_{N_i}=-D_i\left[\nabla n_i + \frac{\hat{\cal S}_in_i}{k_BT}\nabla T
-\frac{q_in_i}{k_BT}E  \right]
\end{equation}
where $D_i=L_{ii}/n_i$ represents the diffusion coefficient, $E=-\nabla V$ is
the local electric field and:
$$
\hat{\cal S}_i=
\overline{\overline {\cal S}}_i-s_i
$$
is the \emph{``Eastman entropy of transfer''}.\cite{eastman}

\subsection{The steady state}
When a temperature gradient is applied to a closed system, after some transient
time, a steady state is reached when the conduction current
(last term in Eq. (5)) exactly balances the 
two other terms coming from diffusion.
In that case the current $J_{N_i}$ of each particle species vanish:
$J_{Ni}=0$, and we can write:
$$
\sum_{i=1}^\nu -\frac{q_i}{D_i}J_{Ni}=0
$$ 
which, taking into account Eq. (5), becomes:
\begin{equation}
\nabla\left(\sum_{i=1}^\nu q_in_i\right)+
\left(\sum_{i=1}^\nu\frac{q_in_i\hat{\cal S}_i}{k_BT}\right)\nabla T-
\left(\sum_{i=1}^\nu\frac{q_i^2n_i}{k_BT}\right) E=0
\end{equation} 
Far from the boundaries we have  charge neutrality and the first term
in the previous equation vanishes. We deduce the local electric field at 
steady state:
$$
E=-\nabla V= S_e\nabla T \qquad {\rm with}
$$
\begin{equation}
S_e=\frac{\sum_{i=1}^\nu q_in_i\hat{\cal S}_i}{\sum_{i=1}^\nu q_i^2n_i}
\end{equation}
corresponding to the Seebeck coefficient.

Substituting this steady electrical field into Equation (5), we have with
$J_{Ni}=0$:
$$
\frac{\nabla n_i}{n_i}=\alpha_i\nabla T \qquad {\rm with}
$$
\begin{equation}
\alpha_i=-\frac{1}{k_BT}\left[\hat{\cal S}_i-
q_i\frac{\sum_{i=1}^\nu q_in_i\hat{\cal S}_i}{\sum_{i=1}^\nu q_i^2n_i}\right]
\end{equation}
defining Soret coefficients.
For simple monovalent electrolytes $A^+B^-$,
 we simply have:
\begin{equation}
S_e=\frac{\hat{\cal S}_{A^+}-\hat{\cal S}_{B^-}}{2e}
\end{equation}
and
\begin{equation}
\alpha_{A^+}=\alpha_{B^-}=-
\frac{\hat{\cal S}_{A^+}+\hat{\cal S}_{B^-}}{2k_BT}
\end{equation}
(The notation used in this paragraph closely follows that
of W\"urger).\cite{wurger}

\subsection{Entropy of transfer}

The \emph{transported entropy} $\overline{\overline{\cal S}}_i$ and the 
\emph{Eastman entropy of transfer} $\hat{\cal S}_i$ 
of most simple ions in aqueous solutions
have been obtained through the measurement of:
\begin{itemize}
\item
The Soret coefficient,
 using e.g conductimetric methods,\cite{turner}
\item
 The ``initial electromotive force''\cite{agarrev}, i.e. the 
difference of potential immediately
 after the temperature gradient is established --well before 
the Soret steady-state has been reached--, in thermocell using
reversible electrodes (Ag/AgCl, Hydrogen electrodes etc...).\cite{ikeda} 
\end{itemize}
For example, from Table I in Ref.~\onlinecite{agar} and  Eq. (9), we
expect, in the infinite dilution limit: $S_e=51\mu$V/K for NaCl,
 $S_e=221\mu$V/K for HCl, and $S_e=371\mu$V/K  for 
TetraButylAmmonium Nitrate in water at 300K.

In the infinite dilution limit, some contribution to the entropy of 
transfer $\hat{\cal S}$ simply arises from the polarization of the surrounding
dielectric medium due to the ion charge. Using hydrodynamic theory and
the Born model\cite{born}, Agar\cite{agarrev,agar} has  estimated 
this contribution to:
\begin{equation}
\hat{\cal S}_0^{Born}=-\frac{(eZ)^2}{16\pi\epsilon_0\epsilon}\frac{1}{R}
\frac{\partial\ln\epsilon}{\partial T}.
\label{dielec}
\end{equation}
 $Z$ is the charge number,  $R$ represents the
ionic radius of the ion, and $\epsilon$ is the relative dielectric constant of
the solvent.
The entropy of transfer of a few simple ions (Na$^+$, F$^-$ ...) 
agrees with this relation. 
For other ions there are also substantial contributions 
 coming from  ion-solvent interactions in the first and second solvation
shells of the ion. Agar\cite{agar}  proposes a simple
classification depending on the overal sign of these additional contributions.
According to the terminology introduced by Gurney\cite{gurney}, an ion is 
called \emph{``structure maker''} if this contribution to the Eastman entropy
of transfer is positive and \emph{``structure breaker''} in the opposite
case. From Table I in Ref~\onlinecite{agar},
while I$^-$, ClO$_4^-$,... ions are \emph{``structure breakers''}, 
most multivalent cations are \emph{``structure makers''}.
A large \emph{``structure making''}
 effect is observed with tetraalkylammonium ions,\cite{agar,petit}
where the measured Eastman entropy of transfer is 10 times higher than the pure
electrostatic contribution expected from Eq.~(\ref{dielec}). 
There is an extensive
literature on \emph{structure making} (or \emph{kosmotrope}) and
\emph{structure breaking} (or \emph{chaotrope}) effects in aqueous 
solutions\cite{marcus1} with different classification criteria. Another
relevant criterium is the change in the hydrogen bonding structure of water
molecules induced by the presence of an ion. The ion is classified as
\emph{structure maker} if it increases the local order, due to hydrogen bonds,
and \emph{structure breaker} in the opposite case. Table 8 in 
Ref.~\onlinecite{marcus1} shows that according to 
this criterium tetrabutylammonium
ions appears also among the ions exhibiting the strongest 
\emph{structure making} effects. Unfortunately the literature concerning ions
in organic solvent is scarce. However we should expect similar 
\emph{structure making} effects in alcohol, which due to the presence of
-OH groups also exhibit local order due to hydrogen bonds.

\section{Experimental}

\subsection{Methodology}
The measurement principle is simple.
A constant temperature gradient is established along the vertical
axis of a cell containing  an electrolyte. 
The ``open circuit'' potential difference $\Delta V$, 
between two points in the liquid, is measured along that temperature gradient, 
using noble metal (platinum) electrodes connected to an electrometer.  We
wait a sufficiently long time (generally more than two hours) such that the
steady state (see Section II B) is fully established. 
At steady state,
we have: $\Delta V=-S_e\Delta T_{Elect}$, where $\Delta T_{Elect}$ 
is the difference of temperature between the two electrodes.

\subsection{ Sample cell for thermoelectric potential measurement}

 The schematic experimental setup is shown in Fig.~1. 
The fluid sample is contained in a hollow cylinder 15 mm high and 10 mm  
diameter machined out from a Teflon$^{TM}$  DuPont parallelepiped. 
The two apertures of the 
cylinder are closed by two horizontal 5 mm thick sapphire windows. In this 
way, the sample that fills the cell is electrically isolated from the
environment. The cell is 
positioned vertically and heated from the 
top by means of a thin film resistance 
glued onto the upper window.
Thanks to the large thermal conductivity of sapphire 
a uniform heating of the sample cross-section is established
 and a stable temperature gradient  settles in. 
The upper window  temperature is controlled within 5 mK stability
by a temperature controller (Lakeshore A340).
 The lower window is maintained at a constant temperature by means 
of a copper pad thermally regulated by a circulating water bath. In our 
experiment, the temperature of the lower window is kept at 
$T_{cold}=(24\pm 0.01 )^o$C and the temperature $T_{hot}$ 
of the upper window is 
varied between 28$^o$C and 75$^o$C. This gives rise to a maximum temperature 
gradient $\approx$34 K.cm$^{-1}$ along the sample length. 
The temperature difference $\Delta T_{Elect}$ between 
the electrodes was measured 
by  type K thermocouples. Figure 2 shows that a linear variation holds
 between $\Delta T_{Elect}$ and $\Delta T =T_{hot}-T_{cold}$.
 This shows that the heat transfer
 along the cell axis is purely conductive and  no convective flow settles 
in the fluid.
The open-circuit potential is measured by two home-made electrodes
 made from a platinum wire ($\phi=$0.33~mm) inserted into a glass tube and 
flame-sealed at the two ends. The tips of the electrodes are flush ground 
and mirror polished with a 3~$\mu$m diamond paste (see inset in Fig. 1).
 The two electrodes are 
located 6~mm apart with their tips positionned along the vertical cell axis.
 Before each run,  the electrodes 
are cleaned for a few hours in a concentrated solution of 33 wt\% HCl and 
rinsed with distilled water.
The cell and the electrodes are  shielded from electromagnetic
radiation by a metal box.
 The open-circuit potential difference
$\Delta V$ between the
 two electrodes is read by a Keithley-6514 electrometer with 
a large input impedance $\approx 2\times 10^{14}~\Omega $. 

\begin{figure}
\includegraphics[width=1.1\linewidth]{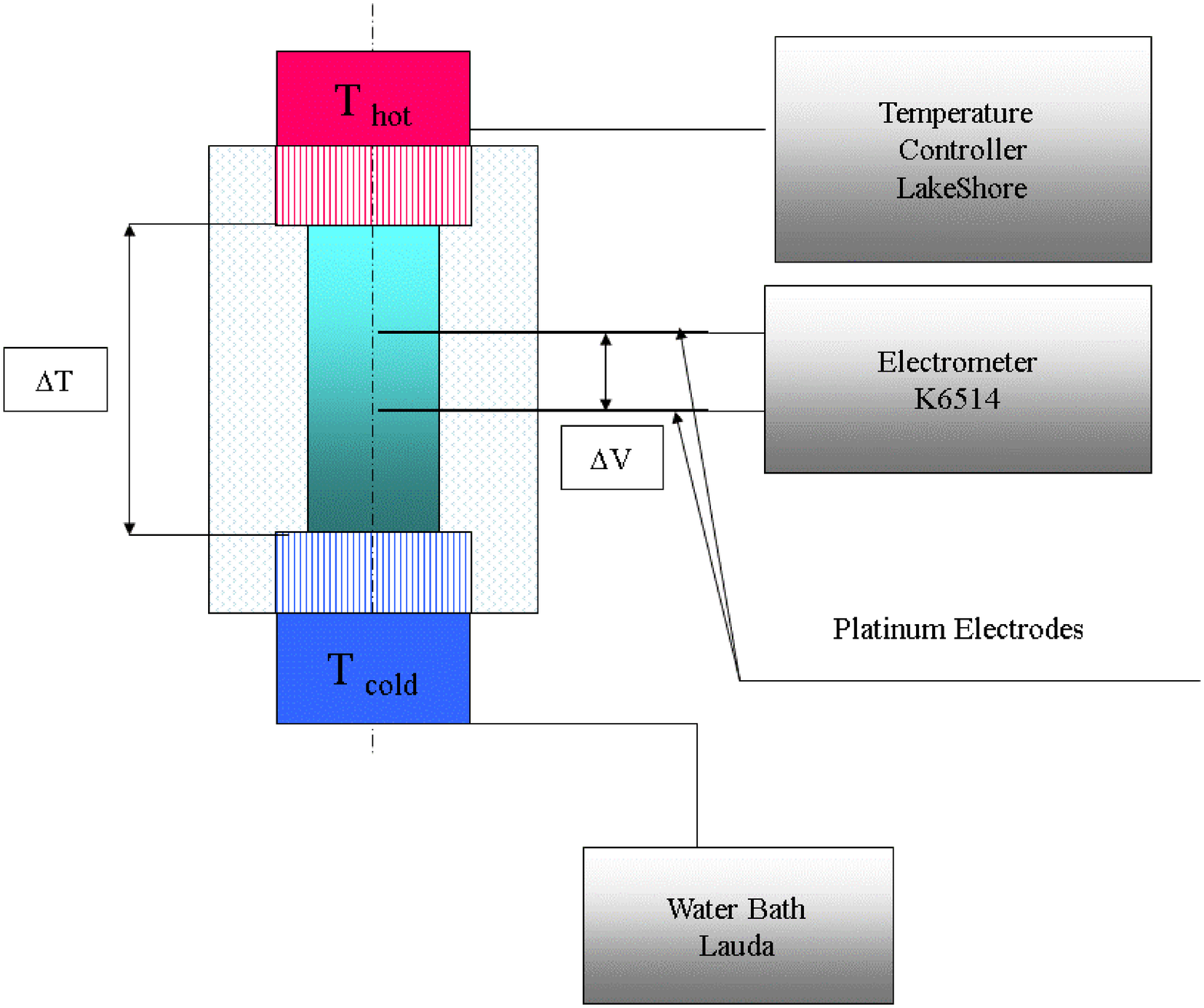}
\includegraphics[width=0.4\linewidth]{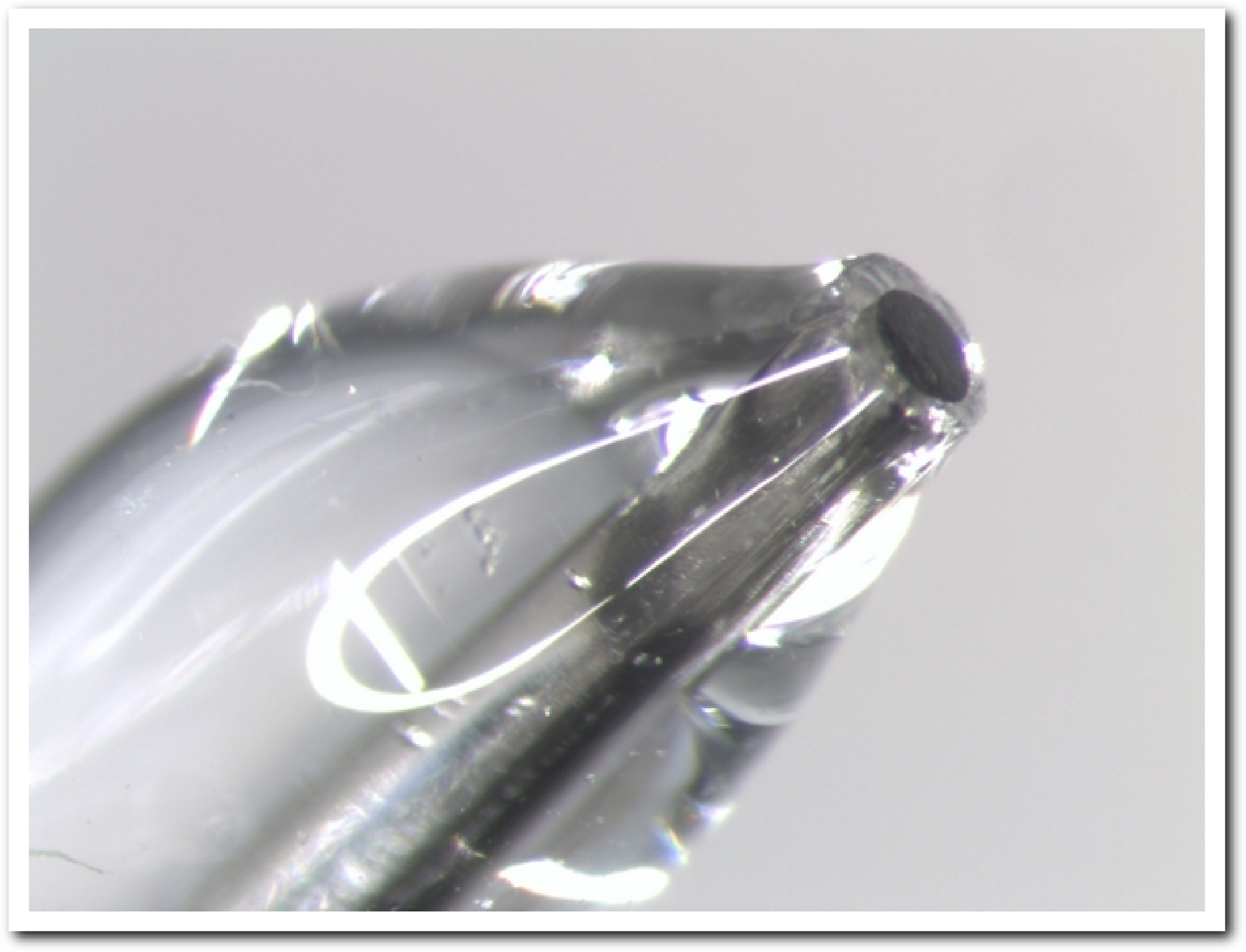}
\caption{\label{figcell} 
 Experimental setup used for the measurement of the thermoelectric
 effect in electrolytes. The cell is made of Teflon. The liquid sample fills 
a vertical cylinder of 10 mm internal diameter and 15 mm height. The 
distance between the two platinum electrodes is 6 mm. The top of the cell 
is heated by a thin-film resistance glued on a sapphire window. The lower
 sapphire window is kept at a constant temperature by a circulating water 
bath. The inset at the bottom left shows the picture of a home-made
platinum electrode.}
\end{figure}

\begin{figure}
\begin{center}
\includegraphics[width=0.9\linewidth]{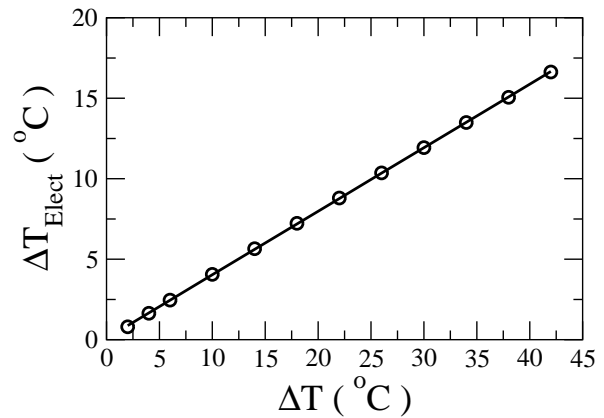}
\end{center}
\caption{\label{etalon}
 Temperature difference $\Delta T_{Elect}$ between the two electrodes
 as a function of the temperature difference $\Delta T$ between the top and 
the bottom of the cell. The cell is filled with 1-dodecanol. The linear
 relation shows that heat transfer is purely conductive. }
\end{figure}
\subsection{Measurement of the thermoelectric potential.}
All  chemicals were purchased from Aldrich and were used without further
 purification. tetrabutyl-ammonium nitrate (TBAN), tetradodecylammonium 
nitrate (TDAN) and tetraoctylphosphonium bromide (TOPB) are of purities
 of 97\%, 99\%, and 97\%, respectively and used as received. 
The purity of 1-octanol and 1-dodecanol is stated to be $>$ 99.5\%.
 Ultrapure water (resistivity: 18.2~M$\Omega$.cm) was obtained from Millipore. 
Mixtures of TBAN in water,
 1-octanol,  1-dodecanol and ethylene-glycol, TDAN in 1-octanol 
and 1-dodecanol, and TOPB 
in 1-octanol were prepared  by weighing. 

The temperature of the heating window is first increased stepwise up 
to $\approx 60^\circ$C by two or four degrees and then decreased back 
to the initial temperature to check reversibility. Typically, a stable 
temperature gradient is established  within half an hour.
 Measurement 
of the open-circuit potential difference $\Delta V$ 
is made in a time interval  lasting
 between one  and a few hours. The values of the Seebeck coefficient 
are obtained from the plateau reached in the potential difference.
 The Seebeck coefficient is computed from the relation: 
$S_e=-\Delta V/\Delta T_{Elect}$,
 with $\Delta T_{Elect}$ deduced from Fig. 2 for each $\Delta T$ step.

\subsection{Electrical conductivity measurements}

The temperature dependent electrical conductivity $\sigma$ is measured in 
a cell with blackened platinum electrodes, and a cell constant 0.80~cm$^{-1}$
(Philips PW 9512/01). 
The in- and out-of-phase components of the impedance
were measured
with an AC LCR auto balancing bridge (HP 4284A) 
in a low frequency range, typically
between 100~Hz and 200~kHz.\cite{marco}
 The reported conductivity values are for a null
out-of-phase component, corresponding therefore to a purely resistive signal.

\section{Results and Dicussion.}

\subsection{Typical measurements}

\begin{figure}[t]
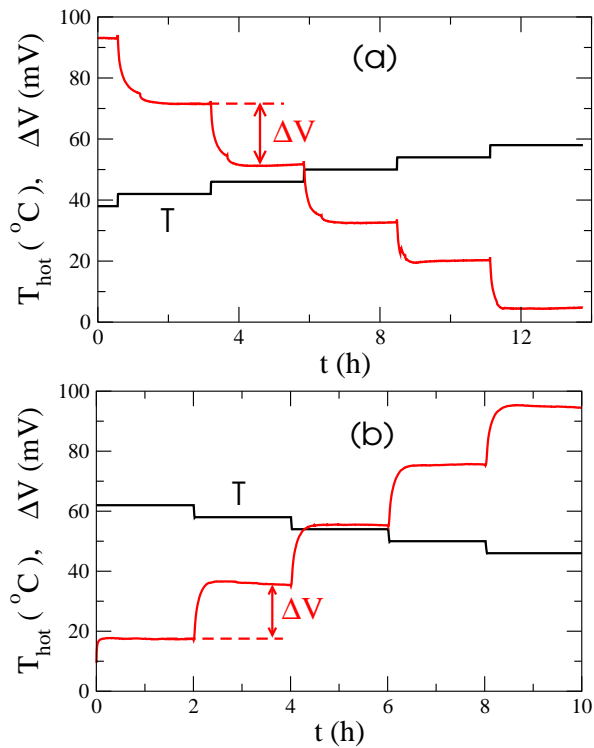

\begin{center}
\includegraphics[width=0.9\linewidth]{fig3a.eps}
\includegraphics[width=0.9\linewidth]{fig3b.eps}
\end{center}
\caption{\label{deltav} 
 Tetradodecylammonium nitrate (0.1M) in 1-octanol: Open-circuit potential
difference $\Delta V$ between the two platinum electrodes when the temperature
$T_{hot}$ of the upper window is changed by steps of : 
(a) 4$^o$C ; (b) -4$^o$C.
The origin for potentials has been arbitrarily shifted for convenience. 
The temperature of the lower window is set to $T_{cold}=24~^o$C.}
\end{figure}

Figure 3-a shows the potential difference $\Delta V$ from the sample TDAN
 in 1-octanol when the temperature $T_{hot}$ of the heating window is changed 
by 4$^\circ $C steps from 38$^\circ $C to 58$^\circ $C. 
It can be noticed that a positive temperature 
change corresponds to a negative $\Delta V$ between the electrodes. 
Conversely, negative (-4$^\circ $C)
 temperature steps give positive $\Delta V$ as it is shown in Fig. 3-b.
 A similar
 behavior was observed with the other  studied electrolytes.
 In Fig.~4, the
 Seebeck coefficient $S_e$ for TDAN in 1-octanol and 1-dodecanol is shown as
 a function of the temperature at half electrode distance.
 It can be observed that the temperature does
 not modify appreciably the Seebeck coefficient within the studied 
temperature range.

\begin{figure}
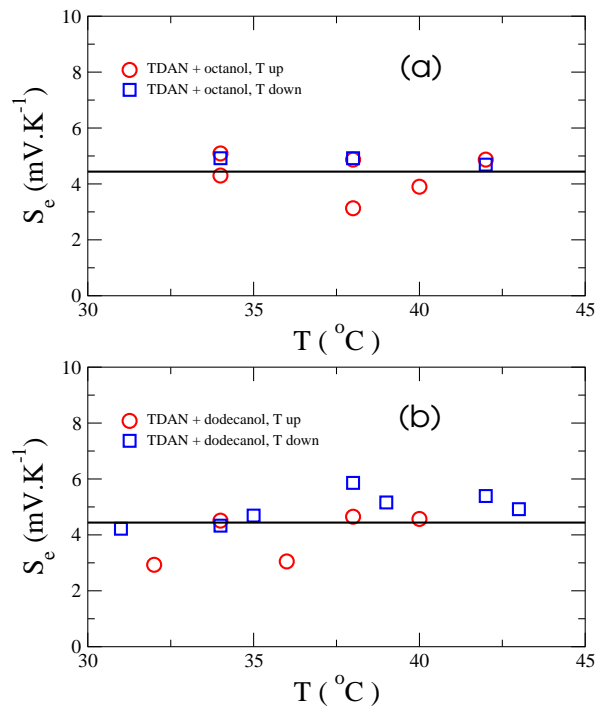

\begin{center}
\includegraphics[width=0.9\linewidth]{fig4a.eps}
\includegraphics[width=0.9\linewidth]{fig4b.eps}
\end{center}
\caption{\label{moyenne}
 Seebeck coefficient $S_e$ as a function of temperature at half electrode
distance.
(a) Tetradodecylammonium nitrate + 1-octanol, (b) Tetradodecylammonium
 nitrate + 1-dodecanol. Circle: increasing temperature; Square: 
decreasing temperature.}
\end{figure}

\subsection{Variations of the Seebeck coefficient with salt concentration}

\color{rouge}
The potential difference $\Delta V$ between the electrodes is measured with
a Keithley-6514 electrometer having 
a large input impedance $\approx 2\times 10^{14}~\Omega $. 
Therefore, the current flowing during the measurement 
of the thermoelectric voltage $\Delta V\approx 10$~mV  is
only of the order of a few hundreds of electrons per second. These electrons 
are provided through reactions at the electrodes. Among the chemical reaction
which can occur at low potentials with platinum electrodes, we might have:
\begin{itemize}
\item
Oxidation of alcohol through carbon adsorption on platinum:\cite{rightmire}

\centerline{Pt + RCH$_2$OH $\to$ Pt-CHOHR + H$^+$+ e$^-$}

\noindent
or through oxygen adsorption on platinum

\centerline{Pt + RCH$_2$OH $\to$ Pt-OCH$_2$R + H$^+$+ e$^-$}

\item
Reduction of nitrate 
oxide:\cite{molodnika}

NO$_3^-$ + 2H$^+$+ 2e$^-$ $\to$ NO$_2^-$ + H$_2$O
\end{itemize}
Reversible adsorption of tetraalkylammonium ions has also
been previously observed at hanging mercury drop electrodes,\cite{sawamoto}
gold electrodes\cite{agnieszka} and CO-coated platinum 
electrodes.\cite{yamakata} The image charge induced in platinum by a positive
tetraalkylammonium ion sticking to the surface could also participate to
electron transport in the external circuit.

Unknown additional thermo-voltages can arise from the entropies 
transported through these reactions.
The following experimental results, showing a linear variation of 
the Seebeck coefficient $S_e$
with the square root of the  tetraalkylammonium-ion concentration, are
indicating that those parasitic thermo-voltages are negligible with respect to
the thermoelectric voltage $\Delta V$ induced by the electric field due to
the density gradient of tetraalkylammonium ion in the cell. 
For TDAN in octanol and TBAN in dodecanol,
the Seebeck coefficient  has been measured in
a restricted range of concentrations $c$
 from 0.001M to 0.65M, due to
solubility limitations. 
Our results are gathered in Table I and
plotted as a function of $\sqrt{c}$ in Fig. 5.
 A linear decrease of the 
Seebeck coefficient with  $\sqrt{c}$ is observed for TDAN in 
octanol. This is in agreement with a Debye-H\"uckel treatment of the 
electrostatic interactions between ions, which predicts a linear variation
of the Eastman entropy  of transport with the square root of the ionic 
strength.\cite{kirkwood}

Using Eq. (9), we expect for tetraalkylammonium nitrate solutions:
\begin{equation}
S_e=\frac{\hat{\cal S}_{TA^+}-\hat{\cal S}_{NO_3^-}}{2e}
\end{equation}
where $\hat{\cal S}_{TA^+}$ is the Eastman entropy of transfer of the 
tetraalkylammonium ion and $\hat{\cal S}_{NO_3^-}$ that of the nitrate ion.
Neglecting the Eastman entropy of transfer of NO$_3^-$ ions 
(in water it is 30 times smaller than that of tetrabutylammonium\cite{agar}),
Eq. (12) reads:
\begin{equation}
S_e\approx \hat{\cal S}_{TA^+}/2e.
\end{equation}
At finite ion-concentation $c$, the Eastman
entropy of transfer varies as:
\begin{equation}
\hat{\cal S}= \hat{\cal S}_0+ \hat{\cal S}'(c)
\end{equation}
where $\hat{\cal S}_0$ is the limit value at infinite dilution. 
$\hat{\cal S}_0$ includes the contribution coming from the polarization
of the surrounding medium given by Eq. (11) within the Born model, plus
short-range interactions with the solvent in the second solvation shell
of the ion. Within Debye-H\"uckel theory, the concentration dependent part
$\hat{\cal S}'(c)$ can be expressed as:\cite{takeyama}
\begin{equation}
 \hat{\cal S}'=\frac{e^2{\cal N}}{4\pi\epsilon_0\epsilon T}\kappa_D
\left[ \frac{1}{12} +\frac{3}{4}\frac{\partial\ln\epsilon}{\partial\ln T}
-\frac{1}{4}\frac{\partial\ln\rho}{\partial\ln T}
\right].
\end{equation}
${\cal N}$ is Avogadro number, $\rho$ is the density of solvent and
$$
\kappa_D=\sqrt{\frac{2e^2 n_i}{\epsilon_0\epsilon k_B T}}
$$
is the inverse Debye length ($n_i= 1000 {\cal N} c$ is the ion density per
$m^3$).

Using for $\epsilon$ and its temperature dependence the values
 from the literature
reported in table II, we find from Eqs. (13-15):
\begin{equation}
S_e=S_e(0)-5.0\sqrt{c} \quad\textrm{(mV/K)}
\end{equation}
for TDAN in octanol at 39$^o$C, and
\begin{equation}
S_e=S_e(0)-7.2\sqrt{c} \quad\textrm{(mV/K)}
\end{equation}
for TBAN in dodecanol at  33$^o$C

Fitting Eqs. (16) and (17) to our results gives $S_e(0)=6.1$mV/K for 
TDAN in octanol and $S_e(0)=8.8$mV/K for  TBAN in dodecanol. The nice
agreement of the calculated slopes with the data (see Fig. 5) 
proves that we are
measuring essentially the entropy of transfer of tetraalkylammonium ions and
that parasitic thermo-voltage contribution due to chemical reactions 
at the electrodes are smaller than  our error bars.

The figure of merit ZT is also plotted in Fig. 5.
 Since the Seebeck coefficient $S_e$ decreases while the ionic
conductivity $\sigma$ increases when the concentration $c$ is raised, there
should be an optimal $c$ corresponding to a maximum $ZT$, expected to
occur at $c\approx 1M$. 
\color{black}

\begin{figure}[]
\begin{center}
\includegraphics[width=0.95\linewidth]{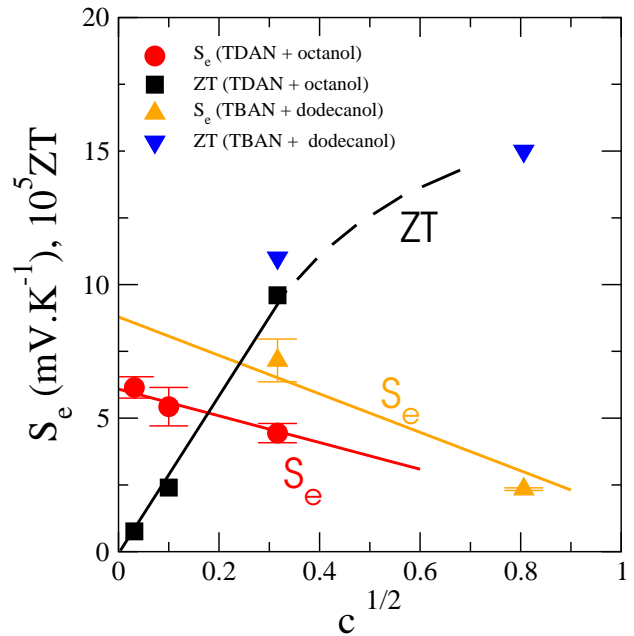}
\end{center}
\caption{\label{sdec}  Seebeck coefficient $S_e$
 as a function of 
the square root of ion concentration $c$(mol/l)
 for TDAN in octanol (red circles) and
TBAN in dodecanol (orange up-triangles). The straight lines are  least square
fit of the data with the slope set to the theoretical value (see text). 
The figure of merit $ZT$ is also
ploted for: TDAN in octanol (black squares) and TBAN in dodecanol
(blue triangle-down). The black dashed curve is a guide to the eyes. 
 }
\end{figure}

\begin{table*}[]
\caption{Seebeck 
coefficient $S_e$, electric conductivity $\sigma$, thermal conductivity 
$\kappa$ of the solvent, and figure of merit $ZT$ as a function of salt
concentration $c$
 for TDAN in octanol and TBAN in docecanol.
$T_{mean}$ is the mean value of the temperature range where $S$ has been
measured. }
\begin{ruledtabular}
\begin{tabular}{ccccccc}
Sample& $c$ & $T_{mean}$ & $S_e$ & $\sigma$ & $\kappa$ & $ZT\times 10^4$ \\
 &(mole/l)& ($^o$C) & (mV/K) & (mS~cm$^{-1}$) & (W~m$^{-1}$~K$^{-1}$) & \\
\hline
TDAN+octanol & $10^{-3}$& 37 & 6.15$\pm$0.4& 0.10 $10^{-2}$ & 0.156\footnote{from Ref. \onlinecite{leneindre}} &0.076\\
TDAN+octanol &  $10^{-2}$&36 & 5.43$\pm$0.7& 0.42 $10^{-2}$ & 0.156$^a$& 0.24\\
TDAN+octanol &  $10^{-1}$&39&4.44$\pm$0.4& 2.42 $10^{-2}$&0.155$^a$&0.96\\
TBAN+dodecanol& $10^{-1}$& 33& 7.16$\pm$0.8&1.17~10$^{-2}$& 0.169$^a$&1.1\\
TBAN+dodecanol& 0.65&37& 2.34$\pm 0.1$& 14.60~$10^{-2}$& 0.169$^a$ &1.5 \\
\end{tabular}
\end{ruledtabular}
\end{table*}

\begin{table*}
\caption{  Seebeck 
coefficient $S_e$, electric conductivity $\sigma$, thermal conductivity 
$\kappa$ of the solvent, 
and figure of merit $ZT$ for various electrolytes at salt concentration
$c=0.1$M. $T_{mean}$ is the mean value of the temperature range where
 $S_e$ has been measured.
The  dielectric constant $\epsilon$ of the solvents with the temperature
derivative of its inverse,
and the ionic radii $R$ of the tetraalkylammonium cations are 
also reported.}
\begin{ruledtabular}
\begin{tabular}{ccccccccc}
Sample& $T_{mean}$ & $S_e$ & $\sigma$ & $\kappa$ & $ZT\times 10^3$&$\epsilon$&
$-\frac{1}{\epsilon^2}\left.\frac{\partial \epsilon}{\partial T}\right|_{T_{mean}}$
&$R$\\
(c=0.1M) & (C) & (mV/K)&(mS~cm$^{-1}$)&(W~m$^{-1}$~cm$^{-1}$)&
 & & (K$^{-1}$) & (nm)\\
\hline
TBAN+dodecanol& 33& 7.16$\pm$0.8&1.17~10$^{-2}$& 0.169\footnote{from Ref. \onlinecite{leneindre}}&0.11& 5.72\footnote{from Ref. \onlinecite{landolt}} &11.69 10$^{-4}$ &0.496\footnote{from Ref. \onlinecite{radius}}\\
TBAN+octanol & 35 & 2.80$\pm$0.3&4.73~10$^{-2}$&0.156$^a$&0.073 & 8.89$^b$
&  9.68 $10^{-4}$ &0.496$^c$\\
TBAN+Ethylene-Glycol&37&3.40$\pm$0.2&60.0~10$^{-2}$&0.256$^a$&0.84&35.5\footnote{from Ref. \onlinecite{akerlof}}&1.38~10$^{-4}$&0.496$^c$\\
TBAN+water& 33& 1.03$\pm$0.1&7.0&0.619$^a$&0.37& 75.29\footnote{from Ref. \onlinecite{fernandez}} & 0.61 10$^{-4}$& 
0.496$^c$ \\
TDAN+dodecanol& 37& 4.44$\pm$0.8&0.66~10$^{-2}$&0.169$^a$&0.024& 5.56$^b$& 12.41 10$^{-4}$&0.694$^c$ \\
TDAN+octanol& 39& 4.44$\pm$0.4&2.42~10$^{-2}$&0.155$^a$&0.096& 8.61$^b$ & 9.98 10$^{-4}$& 0.694$^c$ \\
TOPB+octanol& 39& 2.84$\pm$0.5&3.02~10$^{-2}$&0.155$^a$&0.049& 8.61$^b$&  9.98 10$^{-4}$& 0.613$^c$\\
\end{tabular}
\end{ruledtabular}
\end{table*}

\subsection{Results at salt concentration $c$~=~0.1M}

We have chosen here to compare the Seebeck coefficient of 7 different 
electrolytes at the same concentration $c=0.1M$, which is below the 
 solubility limit of all the salts studied here.

Table II gives the averaged value of the Seebeck coefficient $S_e$ measured in
 the [30$^o$C - 45$^o$C] range for both positive 
and negative temperature variation. 
The Seebeck coefficient varies from $\approx$2.8 mV.K$^{-1}$ 
 to $\approx$7 mV.K$^{-1}$ among different electrolytes, the latter
 value being obtained with TBAN in 1-dodecanol. The averaged dielectric
 constant $\epsilon$, the temperature derivative of its inverse
$(-1/\epsilon^2)d\epsilon/dT$ for pure 1-octanol and 1-dodecanol, 
and the ionic radius $R$ of tetraalkylammonium
cations  are also reported in Table II. 
To compare to the contribution of the pure bulk dielectric effect, calculated 
according to Born approximation   [Eq. (11)], we have plotted in Fig. 6
 all our results at 0.1M as a function of $(1/R\epsilon^2)(-d\epsilon /dT)$.
%
The Born model, Eq. (11), which only takes
into account the interaction of the charge carried by an ion with the
surrounding dielectric medium (assumed  uniform) would lead to:
\begin{equation}
S_e\approx \frac{\hat{\cal S}_{TA^+}}{2e}
=-\frac{e}{32\pi\epsilon_0\epsilon^2R}\frac{d\epsilon}{dT}.
\end{equation}
This gives a slope $a_{Born}=180~$mV.nm (straight line in Fig. 6).

For all data, the pure electrostatic contribution represents only one tenth
of the measured Seebeck coefficient. We infer that the main contribution to
the Seebeck coefficient (i.e. to the Eastman entropy of transfer) originates
mainly from kosmotrope or \emph{structure making} effects. 
Kosmotrope and chaotrope
effects have been extensively studied in aqueous solution.\cite{marcus1} 
Much less theoretical and experimental work has been accomplished in
non-aqueous electrolyte, in particular in alcohols in which hydrogen bonding
is also important.\cite{marcus2}
 We hope that the large kosmotrope effect that we point out
for tetraalkylammonium ions in alcohols will encourage further theoretical
and experimental investigations.

%
\color{rouge}
Comparing the results obtained with tetrabuthylammonium in four solvents
(red circles in Fig.~6), we observe an increase of the Seebeck coefficient
which roughly scales with
the temperature derivative of the inverse dielectric constant:
$-(1/\epsilon^2)d\epsilon /dt$. There is for instance an increase by a factor
20 of this quantity from water to dodecanol (see Table II) and the 
Seebeck coefficient is one order of magnitude higher in dodecanol, compared 
to water solvent. This is not surprising since the kosmotrope effects 
(dipolar forces, hydrogen bonding) which characterize the second solvation
shell are mainly govern by effective electrostatic forces.

In the same solvent, dodecanol, the two results
for TBAN  and TDAN ( red and blue circles in Fig.~6)
 appear to be roughly proportional to $(1/R)$. However concerning TBAB, TDAN
and TOPB in octanol (red, blue, magenta squares) the dependence on $R$ is less
conclusive.
\color{black}
%

\begin{figure}[!h]
\begin{center}
\includegraphics[width=1.\linewidth]{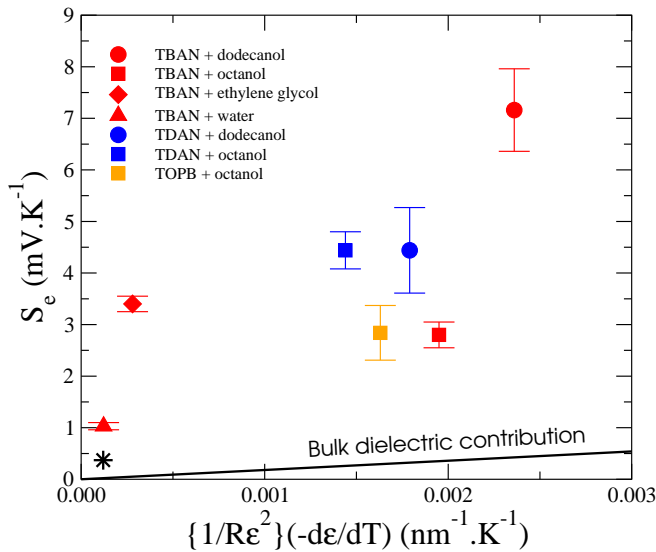}
\end{center}
\caption{\label{all} 
 Seebeck coefficient of different tetraalkylammonium salts, at $c$=0.1M,
in various solvents, as a function of $(1/R\epsilon^2)(-d\epsilon/dT)$.
We have also compared the Seebeck coefficient that we
 measured for TBAN in water (up-triangle)
to that, (at infinite dilution) calculated
from Table I in Ref. \onlinecite{agar}, using Eq. (9) (star). The results
agree within the average error bars.
The color code is red for TBAN, blue for TDAN and orange for TOPB. Different
symbols are attributed to each solvent: circles for dodecanol, squares for
octanol and diamond for ethylene glycol.
 The straight line represents the pure
electrostatic contribution arising from the polarisation of the dielectric
medium (assumed as uniform) by the ion charge. It is one order of magnitude
lower. This suggests a strong kosmotrope effect of tetraalkylammonium ions
in alcohols.
 }
\end{figure}

\section{Conclusion and perspectives}

We have obtained remarkably high values of the Seebeck coefficient, up to
$S_e=7$~mV/K, for tetrabutylammonium nitrate in dodecanol. 
However the highest figure of merit  reached in this
study is of the order of $10^{-3}$, i.e. 10 times lower than those obtained in
aqueous potassium ferrocyanide/ferricyanide solutions.\cite{nano}
It should be emphasized that
 a thorough optimization of the concentration $c$ to obtain 
the largest figure of merit has not been conducted.
 We are now working at
further optimizations which could provide  efficiencies comparable  
 to the aqueous ferrocyanide/ferricyanide solutions.
 The electrical conductivity $\sigma$ increases and the 
Seebeck  coefficient $S_e$ decreases with increasing $c$, and this 
would lead to a maximum of $ZT\approx S_e^2\sigma$ at $c\approx$1M.

In order
to reach performances comparable to those of solid-state devices,
 both the Seebeck coefficient and electric conductivity need to be increased
by a factor of 10, which represents a challenging but not unfeasible goal. 

Having electrolytes, in liquid state or embedded in gels, with Seebeck 
coefficient of the order of 10 to 100~mV will present a technological
breakthrough in 
the perspective of cost-effective low-grade energy harvesting for two
principal reasons:
i) These materials are cheap and
abundant compared to thermoelectric semi-conductors.
ii)
In order to reach voltages of a few Volts, solid-state devices
with $S_e\approx 1mV/K$ and $\Delta T\approx 20~^o$C require an assembly of 
hundreds of P type and N type thermoelectric
elements in series. Having materials with a Seebeck coefficient of order
10-100~mV/K would lower this number down to  few units.  

\section*{Acknowledgments}
We thank  S. Auma\^itre, C. Bataillon, J. Boisson
 for enlightening discussions and C. Gasquet-Wiertel
for technical support.
This project was partly funded by the CEA \emph{``Nanoscience''} program and by
CEA/DSM program \emph{``Energie Bas Carbone''}.

\section*{references}

\end{document}